\documentstyle[12pt]{article}
\begin{document}
\def\st{\scriptstyle}
\def\sst{\scriptscriptstyle}
\def\mco{\multicolumn}
\def\epp{\epsilon^{\prime}}
\def\vep{\varepsilon}
\def\ra{\rightarrow}
\def\ppg{\pi^+\pi^-\gamma}
\def\vp{{\bf p}}
\def\ko{K^0}
\def\kb{\bar{K^0}}
\def\al{\alpha}
\def\ab{\bar{\alpha}}
\def\be{\begin{equation}}
\def\ee{\end{equation}}
\def\bea{\begin{eqnarray}}
\def\eea{\end{eqnarray}}
\def\CPbar{\hbox{{\rm CP}\hskip-1.80em{/}}}
\def\K{kagom\'e }
\def\s{\sigma}
\def\t{\tau}
\def\d{\delta}

\def\a{\alpha}
\def\bs{\bigskip}
\def\cl{\centerline}
\def\ra{\rightarrow}
\def\a{\alpha}
\def\b{\beta}
\def\g{\gamma}
\def\d{\delta}
\def\l{\lambda}
\def\bu{{\bf u}}
\def\bv{{\bf v}}
\def\bw{{\bf w}}
\def\p{\phi}

\def\L{${\cal L}$ }
\def\LL{${\cal L}'$ }
\def\si{\sigma}
\def\ur{\nearrow\ }
\def\ul{\nwarrow\ }
\def\dr{\searrow\ }
\def\dl{\swarrow\ }
\def\up{\uparrow\ }
\def\both{\leftrightarrow\ }

\def\be{\begin{equation}}
\def\ee{\end{equation}}
\def\bea{\begin{eqnarray}}
\def\eea{\end{eqnarray}}

 \begin{center}
{\Large\bf Two exactly soluble lattice models in three dimensions}\\
 
\bigskip
{F.Y. Wu}\\
 
\bigskip
{Department of Physics, Northeastern University}\\

{Boston, Massachusetts 02115}\\
 \end{center}

\vskip 2cm
   As a prelude to what might be  expected as forthcoming breakthroughs
in finding new approaches toward solving   three-dimensional
lattice models in the twenty-first century, we
 review the exact  solutions of two
lattice models in three dimensions obtained using 
the    conventional   combinatorial
and  transfer matrix approaches.

 \newpage

\section{Introduction} 
 The twenties  century has witnessed the birth and   growth of
the field of exactly solvable models in statistical mechanics.
It began with the   Lenz-Ising model of Ising\cite{ising} and its
solution in two dimensions by Onsager.\cite{onsager}
   Beginning in the 1960's the field of exactly solvable models
grew phenomenally to encompass vertex-type models\cite{lieb,liebwu}
 and related problems.\cite{baxter}
But most of the new soluble models  are limited to  two dimensions.
To be sure, 
several three-dimensional 
systems have been studied with various degrees of success in recent years.
They include the Zamolodchikov model$\>$\cite{zam} solved by Baxter,\cite{bax}
its more recent extension by Bazhanov and Baxter,\cite{bb}
and  two models solved by this author and collaborators.\cite{wu,hw,hw1}
But these  solutions invariably suffer defects of one kind or
another, and the prevailing consensus
is that  the solution of genuine three-dimensional systems must await for
major breakthroughs.  Before the arrival of the new breakthroughs,  however, it is
instructive to pause, and examine the usefulness
of  conventional approaches when applied to  three-dimensional systems.
  
There are generally two conventional approaches,  
namely, 
a combinatorial approach 
using graphical methods including  Pfaffians and an
 algebraic approach using the method of the transfer matrix,
both of which
   have  been applied with 
some success in treating
three-dimensional systems.
Here, we  review two such  solutions.\cite{wu,hw}
  It is hoped that, with an  understanding of what has been 
  done and what cannot be done with the conventional approaches, one can then proceed 
ahead to explore new lines of inquiry toward  solving genuine
three-dimensional lattice models  in the twenty-first
century.

\section{The combinatorial solution of an $O(-1)$ model}
  \subsection{Definition of the model}
Consider a simple cubic lattice ${\cal L}$ of ${\cal N}=N_1\times
N_2 \times N_3$ sites.
The partition function of an $O(n)$ model$\>$\cite{nien,nien1}
 on ${\cal L}$ is the graph generating function
\be
Z(n)=\sum_{\rm closed \>polygons} n^\ell z^b, \label{graph}
\ee
where the summation is taken over all closed  non-intersecting polygonal
configurations on ${\cal L}$, $\ell$ is the number of polygons, and $b$ is the
number of edges contained each configuration.
We shall restrict to polygons which 
   are directed in a generally negative
 to   positive direction along the three principal axes,
hence a special three-dimensional system.
Assuming periodic boundary conditions,  the lines
will  then loop around the lattice, with each loop 
cutting  one or more times a plane  perpendicular to the $(1,1,1)$
direction. While the evaluation of (\ref{graph}) for general $n$ is still open,
we consider here the $n=-1$ problem and are interested in the per-site partition
function, or ``free energy",
\be
f= \lim_{{\cal N}\to \infty} {\cal N}^{-1} \ln Z(-1).
\ee
 
\subsection{Equivalence with a dimer problem} 
Our strategy is to establish a one-one
correspondence between the configurations in (\ref{graph})
and dimer configurations on a related lattice ${\cal L}'$,
and that the latter can be  evaluated as a Pfaffian.
To construct ${\cal L}'$,  we 
 split each
lattice point of ${\cal L}$ into two, with 
one  point  attached to the  three positive
axes and the other to the three negative axes as shown in Fig. 1. 
The 
dimer lattice \LL is then obtained   by  reconnecting  the
split pairs with  new inserted edges.  We shall associate
dimer weights  $z$ to edges originally in ${\cal L}$, and weights
1 to the newly inserted edges as shown.

The
one-one correspondence between vertex configurations on \L and
dimer configurations on ${\cal L}'$ can be seen by
superimposing any given dimer configuration $C_i$ with
a standard one,  $C_0$, in which 
  all   inserted  edges are covered by dimers.
The superposition  of the two dimer configurations produces 
a graph of transition cycles,\cite{kast} or polygons,
  belonging to  one of two kinds: i) double dimers placed on
new  edges, and ii) line configurations looping around the lattice. 
In the latter case the line configurations form precisely the same polygonal 
 configurations  as in (\ref{graph}).  Conversely, to each  
polygonal configuration in (\ref{graph}),
there exists a unique dimer configuration $C_i$ which, when superimposed 
with $C_0$, produces the line configuration in question.
This establishes the desired correspondence.

Now there are ${\cal N}$ sites on  ${\cal L}$, and 
this leads us to consider a Pfaffian ${\rm Pf} A$ of dimension $2{\cal N}$.
The convention of fixing  signs to lattice edges is to 
direct edges of \LL such that 
the  element $A_{\alpha\beta}$
is positive (negative) if an edge is directed
from site $\alpha$ to $\beta$ ($\beta$  to $\a$).  
We choose to
direct
all edges of
\LL  
along the positive directions as shown in Fig. 1.   Consider a typical term in the 
Pfaffian corresponding
to a dimer configuration $C_i$.  The sign of this term 
relative to the term corresponding to $C_0$ is
the product of the signs of the transition cycles produced by
the superimposition of $C_i$ and $C_0$.  The rule is that
each transition cycle carries a sign $(-1)^{m+1}$, where $m$
is the number of arrows pointing in a given direction when
the transition cycle is traversed.\cite{kast}  For transition cycles consisting of 
double dimers we have $m=1$ so that the sign is always positive.
For other transition cycles, or loops, we have always
$m={\rm even}$, so that each loop carries a factor $-1$.
As a consequence, the relative sign of the term 
is $(-1)^\ell$ where $\ell$ is the number of loops. 
This establishes that,
with our choice of signs for $A_{\alpha\beta}$,
the Pfaffian $A$ is precisely the dimer generating function (2) with $n=-1$.

\subsection{The free energy}  
The Pfaffian ${\rm Pf} A$ is 
 the square root 
of a $2{\cal N}\times 2{\cal N}$
antisymmetric matrix $A$.
 Explicitly, we have
\be
Z(-1) = {\rm Pf}\> A =\sqrt {\det A}
\ee
where 
\bea
A&=&
\pmatrix{ 0 &  1  \cr
		    -1 & 0  \cr}
 \otimes I_{N_1}\otimes I_{N_2}\otimes I_{N_3} 
+\pmatrix{    0     & z \cr
		  -z &0 \cr}
\otimes \biggl[  H_{N_1}\otimes I_{N_2}\otimes I_{N_3}\nonumber \\
 && \hskip 2.5cm  
	      + I_{N_1}\otimes H_{N_2}\otimes I_{N_3}
	      +  I_{N_1}\otimes I_{N_2}\otimes H_{N_3}\biggr]. \label{a2}
\eea
Here, $I_N$ is the $N\times N$ identity matrix,
 $H_N$ is an $N\times N$  matrix
given by
\be
H_N =  \pmatrix{0 & 1 &        0 & \ldots & 0\cr
		   0 & 0 &1 & \ldots  & 0\cr
		       \vdots   &\vdots      &      \vdots&\ddots &\vdots \cr
			0  &    0    &    0    & \ldots  & 1 \cr
			  -1&    0     &   0     &  \ldots     & 0 \cr}. 
			  \ee
  Now, $H_N$  can be diagonalized  with 
 eigenvalues $e^{i\p}$, where
$\p = 2\pi n/N$, $n=1,2,\cdots,N$. Replacing the $H$ matrices by diagonal ones
with its eigenvalues in the diagonal, one obtains 
\bea
Z(-1) 
&=&\prod_{n_1=1}^{N_1}\prod_{n_1=1}^{N_2}\prod_{n_3=1}^{N_3}
\biggl[{\rm det} \pmatrix{0 & 1+\sum_i z e^{2\pi i n_i/N_i}\cr
-1-\sum_i z e^{-2\pi i n_i/N_i} & 0 \cr}\biggr]^{1/2} \nonumber \cr
&=& \prod_{n_1=1}^{N_1}\prod_{n_2=1}^{N_2} \prod_{n_3=1}^{N_3}
\left\vert 1+\sum_{i=1}^3 z e^{2\pi i n_i/N_i} \right\vert.
\eea
This leads to the thermodynamic limit of the free energy
\be
 f =
  {1\over {(2\pi)^3} }\int_0^{2\pi} d\theta_1 \int_0^{2\pi} d\theta_2
\int_0^{2\pi} d\theta_3  \ln \left\vert 1+\sum_{i=1}^3 z e^{i\theta_i} 
\right\vert. \label{freeenergy}
\ee
  
\subsection{The critical behavior}
The analysis of the critical behavior of the free energy 
(\ref{freeenergy}) is facilitated
by the use of the integration formula  
\be
 {1\over {2\pi}} \int_0^{2\pi} d\theta \ln \vert A+Be^{i\theta}\vert
  =\ln {\rm max} \{|A|, |B|\} \label{int}
\ee
which holds for any complex $A$ and $B$.
We realize bond weights 
by introducing bond energy $\epsilon$, and write
without the loss of generality
$z = e^{-\beta \epsilon}<1,$
where $\beta=1/kT$. 
 Then, multiplying the quantity inside the absolute signs 
in (\ref{freeenergy}) by $e^{-i\theta_1}$
and    
carrying out the  integration over $\theta_1$,
 we obtain 
\bea
f &=& 0, \hskip 6.7cm T<T_c \nonumber \\
&= &{1\over {(2\pi)^{2}} }\int d\theta_2\cdots \int_{\cal R} 
  d\theta_3
 \ln \left\vert z(1+\sum_{j=2}^d  e^{i\theta_j}) 
\right\vert,\hskip 0.5cm T>T_c,
\eea
where  $T_c$ is
defined by
\be
3z =1,\label{tc}
\ee
 and the integrations are taken
over the regime ${\cal R}$ specified by $z\vert 1+ e^{i\theta_2}+ e^{i\theta_3} 
\vert>1$. 
Clearly, the free energy is non-analytic in $T$ at $T_c$  with the system
 frozen  below $T_c$. 
  
To find out the nature of the transition, we evaluate
the energy
\begin{eqnarray}
U&=&-{\partial \over {\partial \beta}} f(-1) \nonumber \\
   &=&
 {\epsilon\over {(2\pi)^2} }
  \biggl( z{\partial \over {\partial z}} \biggr)
\int_0^{2\pi} d\theta_1 
 \int_0^{2\pi} d\theta_2
\ln {\rm max} \{ z, \vert x
\vert \} \nonumber \\
&= & 
{\epsilon\over {(2\pi)^{2}} }   \int d\theta_1
 \int_{\cal R} d\theta_2  \label{u}
\end{eqnarray}
 where 
\be
x=1+ z(e^{i\theta_1}+e^{i\theta_2}),
\ee
and ${\cal R}$ is the regime $z>\vert x \vert$.
In writing out the second line in (\ref{u}), we have  carried out
the integration over $\theta_3$.  The last line in (\ref{u}) then follows from
the fact that
 $x$ is independent of $z$  
so that the regime $\vert
x\vert> z$ gives rise to no contribution after taking the derivative.

It is clear that the volume of ${\cal R}$ is small near $T_c$.  
A detailed analysis\cite{hw} shows that 
\be
U\sim |T-T_c|
\ee
 and hence the critical exponent
$\alpha = 0.$
Explicitly, one finds
 \bea
 U&=& 0,\hskip 6.5cm T<T_c  \nonumber \\
&=& {{2\epsilon }\over {\pi^2}} \int_{z -1}^{2z} 
{{du}\over \sqrt{4z^2-u^2}}\cos^{-1} \biggl({{1-u^2-z^2}\over {2uz}}\biggr),
\hskip 0.5cm T>T_c 
\eea
from which one obtains
\be
c_v = {{dU}\over {dT}} ={3\sqrt 3\over {2\pi}} (\ln 3)^2 k,
\hskip 1cm T\rightarrow T_c+.
\ee
Thus, the specific heat has a cusp singularity at $T_c+$.

The analysis of this section can be extended to the $O(-1)$ model 
defined by (\ref{graph})
in $d$
dimensions in which continuous lines run in a preferred direction
along all $d$ principal axes of a $d$-dimensional hypercubic lattice.\cite{wu}
In this case one again finds the system frozen  below $T_c$, but with $T_c$
 defined by $dz=1$. One also finds the critical exponent $\alpha = (d-3)/2$.

\section{The transfer matrix solution of an interacting layered  dimer system}
 \subsection{Definition of the model}
Consider  
 a three-dimensional model
consisting of $K$  layers  of honeycomb dimer
lattices.  The dimers, which carry weights $u, v, w$  along the
three honeycomb0 edge directions, are  close-packed within each layer 
and, in addition, interact between layers.
The situation is shown in Fig. 2.
For two dimers incident at the  same 
  site
in  adjacent layers, the interaction energy is given in Table 1.

The two-dimensional honeycomb dimer system can be formulated
as a five-vertex model, namely, 
an ice-rule model with the weights\cite{wu68}
\be
\{\omega_1, \omega_2, \omega_3, \omega_4, \omega_5, \omega_6 \}
=\{ 0,w,v,u,\sqrt{uv},\sqrt{uv} \}. \label{5vweight}
\ee 
The 5-vertex model is defined on a square lattice  of
size $M\times N$  mapping  to an honeycomb lattice of $ 2MN$
sites.\cite{wu68}  The mapping is such that
the edge state $\alpha = +1$ ($\beta=+1$) corresponds to the presence
of  a $v$ ($u$) dimer.

\begin{table}[t]
\caption{Dimer interacting energy between two dimers incident at the same site 
of adjacent layers.}
 \vspace{0.4cm}
\begin{center}
\begin{tabular}{|c|c|c|l|}
\hline
 layer $k\to k+1 $ & $w$ & $v$ & $u$ \\ \hline
$w$ & $0$ & $-2h/3$ & $\>\>\>2h/3$ \\ \hline
$v$ & $\>\>\>2h/3$ & $0$ & $-2h/3$ \\ \hline
$u$ & $-2h/3$ & $\>\>\>2 h/3 $ & $\>\>\>\>\>\>0$ \\ \hline
\end{tabular}
\end{center}
\end{table}

The $K$ layers of $M\times N$ square lattice form a
a simple-cubic lattice ${\cal L}$ of size ${\cal N}= K\times M \times N$
with periodic boundary conditions.
  Label sites of ${\cal L}$ by indices $\{m,j,k\}$,  with $1\leq m\leq M$,
$1\leq j \leq N$ and $1\leq k \leq K$.
   Label the state of the
 horizonal (vertical) edge 
incident at the site $\{m,j,k\}$ in the direction of, say, decreasing $\{m,j\}$
 by  $\alpha_{mjk}$ ($\beta_{mjk}$).
 Denote the 
vertex weight at site $\{m,j,k\}$ 
by $W_{mjk}$ and the interacting Boltzmann factor between the $\{m,j\}$ sites
 of two adjacent layers $k$ and $k+1$  
by $B_{mjk}$.
 It can then be verified that the
interaction given in Table 1 can be written 
in the form of 
   \begin{equation}
B_{mjk} 
  = {\rm exp}\biggl(h \bigl( \alpha_{j} \tilde \beta_{j} -
\tilde \alpha_{j+1} \beta_{ j}' \bigr) \biggr)  ,
 \label{inter}
\end{equation}
where for
 convenience we have suppressed the
subscripts $m$ and $k$ and introduced the notation
\begin{equation}
\beta_{m+1,j,k} \rightarrow  \beta_j', \hskip 1cm \beta_{m,j,k+1}\rightarrow
\tilde \beta_j, \label{convention}
\end{equation}
and similarly for the $\alpha$'s. 
Note that as a consequence of the ice rule  
 the quantity 
\begin{equation}
y_k = {1\over N}\sum_{j=1}^N \beta_{j}=
{1\over N}\sum_{j=1}^N \beta'_{j} \label{conserved}
\end{equation}
 is conserved within each layer and is independent of $m$. 
 
The problem at hand now is the evaluation of the
partition function
\begin{equation}
Z_{MNK} = 
\sum_{\alpha_{mjk}} \sum_{\beta_{mjk}}
\prod_{k=1}^K \prod_{m=1}^M \prod_{j=1}^N \biggl(
B_{mjk} W_{mjk} \biggr) \label{partition} 
\end{equation}  
where the summations  are taken over all edge states 
$\alpha_{mjk}$ and $\beta_{mjk}$,
and the per-site ``free energy"
\begin{equation}
f= \lim _{M, N, K\to \infty} (MNK)^{-1} \ln Z_{MNK}. \label{free}
\end{equation}

\subsection{The transfer matrix}
 The partition function (\ref{partition}) can be evaluated  by applying a transfer
matrix in the vertical direction. 
In a horizontal cross section of ${\cal L}$
 there are $NK$ vertical edges.   Let
$\{\beta_m\} = \big\{\beta_{mjk} \big| 1\leq  j \leq N, 1\leq k \leq K \bigr\}$,
$1\leq m\leq M $
denote the states  of these $NK$ vertical edges,
and define a $2^{NK} \times 2^{NK}$  matrix {\bf T} with elements
\begin{equation}
T(\{\beta_m\}, \{\beta_{m+1}\}) = \sum_{\alpha_{mjk}}
\prod_{k=1}^K \prod_{j=1}^N (
B_{mjk} W_{mjk} ). \label{transfer}
\end{equation}
 Then one has
\be
Z_{MNK} = \sum_{\beta_{mjk}} \prod_{m=1}^M
 T(\{\beta_m\}, \{\beta_{m+1}\}) 
 \sim \Lambda_{\rm max}^M, \label{z}
\ee
where $\Lambda_{\rm max}$ is the largest eigenvalue of {\bf T}.
 The validity of the ice rule
within each  layer now permits one to use a global  Bethe ansatz in the analysis.

The interlayer interaction (\ref{inter}) leads to
a considerable simplification of the transfer matrix.
It can be shown\cite{hw1} that we have
 \be
\prod_{m=1}^M \prod_{j=1}^N B_{mjk} = 
\prod_{m=1}^M {\rm exp} \bigg[Nh\bigl(\alpha_{m,1,k}y_{k+1} 
- \alpha_{m,1,k+1} y_k \bigr) \biggr]. \label{b}
\ee
As a result, only the conserved quantities $y_k$
and the state $\alpha_{m,1,k}$ appear in the final product.
This leads us to recast the partition function (\ref{partition}) in the form of
\begin{eqnarray}
Z_{MNK}&=&
  \sum_{\beta_{mjk}}
 \prod_{m=1}^M T^{\rm eff}\bigl( \{\beta_m\}, \{\beta_{m+1}\}
\bigr)   \nonumber \\
&=& {\rm Tr} \bigl({\bf T}^{\rm eff}\bigr)^M \label{partition1}
\end{eqnarray}
 with 
\be
T^{\rm eff}\bigl( \{\beta_m\}, \{\beta_{m+1}\} \bigr)
=\sum_{\alpha_{mjk}}  \prod_{k=1}^K \biggl( 
e^{Nh\alpha_{m,1,k}(y_{k+1}-y_{k-1})}
\prod_{j=1}^N W_{mjk} \biggr). 
 \label{transfer1}
\ee
 The problem is now reduced to one of finding the largest eigenvalue of
the  matrix {\bf T}$^{\rm eff}$ with matrix elements as shown.
  The task is now 
 considerably simpler 
since one needs only to keep track  of  two-dimensional systems.

\subsection{The free energy}
 The eigenvalues of the  effective matrix 
are obtained by applying a global Bethe ansatz 
consisting of the usual Bethe ansatz for each layer. 
The algebra is straightforward and the result is
\begin{equation}
Z_{MNK} \sim \max_{\{ n_k\}} \prod_{k=1}^K \bigl[\Lambda_L (n_k)\bigr]^M , \label{6v}
\end{equation}
with
\be
 \Lambda_L (n_k)=e^{2h(n_{k+1}-n_{k-1})} \omega_4^{N-n_k}
 \prod_{j=1}^{n_k} \biggl[{{\omega_2\omega_4+\omega_5\omega_6 {z_j^{(k)}}}\over 
{\omega_4 }} \biggr] 
\label{evalue0}
\ee
being  the eigenvalue for $\alpha_{m,1,k}=-1$
  and,
 for each $1\leq k \leq K$, the $n_k$ complex numbers 
$z_j^{(k)}$ are 
 \begin{equation}
z_j^{(k)} = e^{i\theta_j} e^{2h(y_{k+1}-y_{k-1})},\>\>\>\>\>
j=1,2,\cdots,n_k \label{zz}
\end{equation}
where $ e^{i\theta_j}$ are  $n_k$ distinct $N$th roots of $(-1)^{n_k+1}$.
 Using (\ref{5vweight}) and (\ref{free}),
this leads  to  the per-site free energy
\begin{eqnarray}
f&= &\ln u + \lim_{K\to\infty}\>\max_{-1\leq y_k \leq 1} {1\over K} \sum_{k=1}^K
   {1\over {2\pi }}
\int_{-\pi (1-y_k)/2}^{ \pi(1-y_k)/2} \nonumber \\
&&\times \ln \biggl(
    {w\over u} + {v\over u} e^{2h(y_{k+1} -y_{k-1})} e^{i\theta} 
     \biggr) d\theta.  \label{free1}
\end{eqnarray}

 \subsection{The phase diagram}
 Analyses of the free  energy (\ref{free1}) lead to the following:\cite{hw1}
For large
$u$, $v$ or $w$, the system is  frozen with complete
ordering of $u$, $v$, or $w$ dimers in all layers, and hence the free energies
\begin{eqnarray}
f_U &=& \ln u, \hskip 2cm  U \>\>{\rm phase} \nonumber \\
f_V  &=& \ln v, \hskip 2cm V \>\>{\rm phase} \nonumber  \\
f_W  &=& \ln w, \hskip 1.9cm W \>\>{\rm phase}. \label{fw}
\end{eqnarray}
 These are   referred 
to as the $U$, $V$, and $W$ phases, respectively.
   For 
large $h$, it is  seen from Table 1 that the 
energetically preferred state is the one in which each 
layer is occupied by one kind of dimers, $u$, $v$, or $w$,
and that the layers are ordered in the sequence of 
$\{u,w,v,u,w,v\cdots\}$. This ordered phase is referred to
as the $H$ phase with the free energy 
 \begin{equation} 
f_H = {1\over 3} \ln \bigl( uvw e^{4h} \bigr), \hskip 1.5cm
  H \>\>{\rm phase}.
\end{equation}
  
More generally one finds that 
  the extremum
value  in (\ref{free1})  always repeats in multiples of 3, 
 namely, with
$$
y_{k+3} = y_k. 
$$
The following extremum sets of $\{y_k\}$ are found:

\medskip
\noindent
1.  $\{y_1, y_2, y_3\} = \{1,1,1\}$:
In this case we have all $y_k=1$, and hence from (\ref{free1})
\begin{equation}
f=f_U,  \hskip 2cm U \>\>{\rm phase}.
\end{equation}
 
\noindent
2. $\{y_1, y_2, y_3\} = \{-1, -1, -1\}$:
In this case we have  all $y_k=-1$, and hence from ({\ref{free1})
\begin{eqnarray}
f &=& \ln u + {1\over \pi} \int_0^\pi \ln \biggl| {w\over u} +{v\over u}
e^{i\theta}
\biggr| d \theta \nonumber \\
&=& f_W, \hskip 1cm w>v \hskip 1cm W \>\>{\rm phase} \nonumber \\
&=& f_V, \hskip 1cm v>w.\hskip 0.95cm V\>\>{\rm phase}
\end{eqnarray}
  
\medskip
\noindent
3. $\{y_1, y_2, y_3\} = \{1, -1, -1\}$:
Substituting this sequence of $y_k$ values into (\ref{free1}) and 
making use of the integration formula (\ref{int})
in the resulting expression, one obtains
\begin{eqnarray}
f&=& \ln u  + {1\over {6\pi}} \int_{-\pi}^\pi \ln \biggl| {w\over u} +
	      {v\over u} e^{2h} e^{i\theta}\biggr| d \theta
  + {1\over {6\pi}} \int_{-\pi}^\pi \ln \biggl| {w\over u} +
	      {v\over u} e^{-2h} e^{i\theta} \biggr| d \theta \nonumber \\
  &=& {1\over 3} f_U + {2\over 3} f_W , \hskip 1.5cm 
ve^{-4h}< ve^{4h}<w \nonumber  \\
  &=& {1\over 3} f_U + {2\over 3} f_V , \hskip 1.5cm w< ve^{-4h}<ve^{4h} \nonumber \\
   &=& f_H .  \hskip 3cm   ve^{-4h} < w <  ve^{4h}. 
\end{eqnarray}
Now the free energies in the second and third lines 
can be discarded since they are always smaller than
the largest of $\{f_U, f_V, f_W\}$.
Thus, this set of
$\{y_k\}$ leads to 
a frozen ordering for  sufficiently large $h$ as indicated in
the last line.

\medskip
\noindent
4. $\{y_1, y_2, y_3\} =\{y, y, y\}$:
In this case all $y_k = y$, 
 $y$  maximizes the  free
energy  (\ref{free1}).  
 Then,  substituting  $y_k=y$ into (\ref{free1}) and 
carrying out a straightforward
differentiation with respect to  $y$, one obtains 
\begin{equation}
f= f_Y(y_0), \hskip 3cm  Y {\rm \>\>phase }
  \label{fy}
\end{equation}
where the extremum $y_0$ is given by
 \begin{equation}
{\pi\over 2} (1-y_0) =\cos^{-1} \biggl[{{u^2-w^2-v^2}\over {2wv}}
\biggr]. \label{y0}
\end{equation}
This is a disorder phase which we
 refer to  as  the $Y$ phase.

\medskip 
\noindent
5. $\{y_1,y_2,y_3\} = \{ y_1, -1,-1\}$:
This is the $H$ phase with the  $u$ layers replaced by 
  layers with $y_k=y_1$,  so that 
the layer ordering is $\{ y_1,w,v, y_1,w,v \cdots\}$.  
This is a partially ordered phase which
we refer to    as the $I_u$ phase.

\medskip 
\noindent
6. $\{y_1,y_2,y_3\} = \{1, y_2, -1\}$:
This is the $H$ phase with the  $w$ layers replaced by 
  layers with $y_{k+1}=y_2$,  so that 
the layer ordering is $\{ u,y_2,v,u, y_2,v \cdots\}$.  
We refer to  this  as the $I_w$ phase.

\medskip
\noindent
7. $\{y_1,y_2,y_3\} = \{1,-1, y_3\}$:
This is the $H$ phase with the  $v$ layers replaced by 
  layers with $y_{k+2}=y_3$,  so that 
the layer ordering is $\{ u,w,y_3,u,w, y_3  \cdots\}$.  
We refer to  this  as the $I_v$ phase.

\medskip  
 Since the phase diagram must reflect the
$\{u,v,w\}$ symmetry of the
 interlayer interaction
given in Table 1,
 it is  convenient to introduce coordinates
\begin{equation}
X=\ln(v/w) \hskip 1cm Y= (\sqrt{3})^{-1} \ln (vw/u^2)
\end{equation}
such that any interchange of the three variables $u$, $v$, and $w$
corresponds to a $120^\circ$ rotation in the $\{X, Y\}$ plane.
The phase boundaries are then determined by equating the free energies
of adjacent phases. The results are collected in Fig. 3.
One finds the following:

\medskip 
$\bullet$ $h<h_0$:
For small $h$ the phase diagram is the same as  that of the $h=0$,
  the diagram shown in Fig. 3(a). 
The phase boundaries between the $\{U,V,W\}$ phases and the $Y$ phase are
 \begin{equation}
u=|v\pm w|.  \label{yuvwboundary}
\end{equation}

$\bullet$ $h_0<h<h_1$:
As $h$ increases from zero, our numerical analyses 
indicate that the $H$ phase appears when $h$
reaches the  value 
 \begin{equation}
h_0 = \frac{3}{8\pi} \int_0^{2\pi/3}
	       \ln ( 2+2\cos \theta ) d\theta =0.2422995....\nonumber
\label{fhfy1}
\end{equation}
  The resulting phase diagram is shown
in Fig. 3(b).    The phase boundary
between the $H$ and $Y$ phases is 
 \begin{equation}
{1\over 3} \ln \bigl( uvw e^{4h} \bigr) = f_Y(y_0).\label{hyboundary}
\end{equation}

$\bullet$ $h_1<h<h_2$:
As $h$ increases from $h_0$, the $I_u, I_v, I_w$ phases appears 
when $h$ reaches a certain value
$h_1$. The resulting phase diagram is shown in Fig. 3(c).
 The numerical value of $h_1$ is
 given by
$$
h_1 = {1\over 4} \ln \biggl({{2v}\over u} \biggr)= 0.2552479\cdots,
$$
where $v/u$ is the   solution of the two equations
\bea
(1+y_0) \ln {v\over u} + {1\over 6} (1+3y_0)\ln 2
& =& {1\over \pi} \int_0^{\pi(1-y_0)/2} \ln(1+\cos\theta)d\theta  \\
 {\pi\over 2} (1-y_0) &=& \cos ^{-1} \biggl({{u^2}\over {2v^2}} -1\biggr).
\eea

$\bullet$ $h_2<h<h_3$: 
As $h$ increases from $h_1$, it was found that the regimes $I_u$, $I_v$, and
$I_w$
extends to infinite when $h$ exceeds the value 
$$
h_2 = (\ln 3) / 4 = 0.2746530\cdots.
$$
  The phase
diagram is shown in Fig. 3(d). 
 
\medskip 
$\bullet$ $h_3<h<h_4$: 
As $h$ increases from $h_2$, it was found that the boundary of the
$H$ phase bulges toward the $U,V,W$ phases along the $30^\circ, 
150^\circ, 270^\circ$ lines, touching the $U,V,W$ boundaries
in these directions when $h$ equals to
$$
h_3 = (\ln 2)/2 = 0.3465735\cdots.
$$
 For $h>h_3$, the $H$ phase borders directly with the 
$U,V,W$ phases with respective boundaries
\begin{equation}
u^2 = vwe^{4h},  v^2 = wue^{4h}, 
w^2 = uve^{4h}. \label{huvwboundary}
\end{equation}
The size of these borders
grows while the $Y$ phase shrinks as $h$ increases.
The phase diagram in this regime is shown in Fig. 3(e).

\medskip 
$\bullet$ $h>h_4$: 
As $h$ increases further from $h_3$, it was found that the 
 $Y$ phase disappears completely when $h$ exceeds the value
$$ 
h_4= {1\over 8 } \ln (1+\gamma^2) = 0.3816955\cdots,
$$ 
where $\gamma$ is the solution of the equation
 \begin{equation}
\gamma -\tan^{-1} \gamma = \pi. \label{gamma}
\end{equation}

\subsection{The critical behavior}

The critical behavior  near all phase
boundaries can be analyzed by expanding the free energy. 
  We find
 all transitions   to be of first order,
except those
between the $\{U,V,W\}$ and $Y$ phases, and between the $\{I_u, I_v, I_w\}$
 and $H$
phases, which are shown below  to be the 
same as that in the 5-vertex model,\cite{wu68,wu67}
namely, a
continuous transition with a square-root singularity in
the specific heat.  This transition,
first reported by this author\cite{wu67}
 in 1967,
is now more widely known as the Pokrovsky-Talapov type
phase transition.\cite{talpol}

To see the occurrence of the Pokrovsky-Talapov type transitions, we note that
 the ordered $U, V, W$ and $H$ phases  (with $y_k=\pm 1$)  have constant
free energies and hence zero
 first derivatives.
 The transition  between the $U,V,W$ phases and
the $Y$ phase  is the same as in two dimensions,
and is continuous.\cite{wu68}
This fact can 
be seen by expanding  the free energy   near $y_0$ as
\begin{eqnarray}
f_Y(y)& =& f_Y(y_0) +(y-y_0)f_Y'(y_0) +{1\over {2!}} (y-y_0)^2f_Y''(y_0) \nonumber \\
&&   +{1\over {3!}} (y-y_0)^3f_Y'''(y_0) +\cdots. \label{taylor}
\end{eqnarray}
Using the expression of $f_Y(y)$ given by (\ref{fy}), one sees that, indeed,
the first derivative $f_Y(y_0)$, $y_0=\pm 1$,  vanishes identically on the boundary
(\ref{yuvwboundary}) which is precisely $f_Y'(y_0) =0$.
  Furthermore, it is also seen that 
$f_Y''(y_0) \sim \sin [\pi(1-y_0)/2] =0$.  Therefore, the extremum of
$f_Y(y)$ given by (\ref{taylor}) occurs at $y=y_{\rm extrm}$ determined from
\begin{equation}
y_{\rm extrm} -  y_0 = \pm
\sqrt{\frac  {2 f_Y'(y_0)}    {-f_Y'''(y_0)}   }
	  \sim t^{1/2},
\end{equation}
where $t= |T-T_c|$,  $T_c$ being the critical temperature.
 Substituting this $y_{\rm extrm}$ into (\ref{taylor}), one obtains
 \begin{eqnarray}
f_Y(y_{\rm extrm})&=&f_Y(y_0) \pm \frac{2}{3} f_Y'(y_0)
	    \sqrt{\frac{2 f_Y'(y_0)}{-f_Y'''(y_0)}}
	   \nonumber \\
	   &=&f_Y(y_0) + c(u,v,w,h) t^{3/2}.
\end{eqnarray}
This leads to a square-root singularity in the specific heat.
  
Applying the same analysis to 
the  $H$ and $I_u$ phases, 
 one concludes
\begin{equation}
f_{I_u}(y_{\rm extrm})
	    =f(y_{10}) + c_1(u,v,w,h) t^{3/2}. \nonumber
\end{equation}
This gives rise again  to a square-root singularity in the specific heat.
 
\section{Summary}
We have described the exact solutions of two lattice models in three dimensions,
an $O(-1)$ model solved   using the combinatorial method of Pfaffians, and
a layered dimer model solved  using an algebraic  transfer matrix.
However, both models stop short of being genuinely three-dimensional.
The $O(-1)$ model suffers two defects.  First, it 
describes line configurations running  only in preferred 
directions.  Secondly, the  Boltzmann weights can be negative.
  The solution of a model without these   restrictions
would help us to solve   the Ising model in three dimensions.
The layered dimer model, while having strictly positive
 weights, describes dimer configurations
in which dimers are confined in  planes.
As a consequence, the critical behavior is essentially two-dimensional.
 
\section*{Acknowledgments}
This  work is supported in part by NSF grant DMR-9614170.

\vskip .5in

\begin{center}
{\bf Figure captions}
\end{center}

\bigskip
\noindent
Fig. 1. (a) A lattice point of the simple cubic lattice ${\cal L}$.
(b) The split of the lattice point in (a) into two points.

\medskip
\noindent
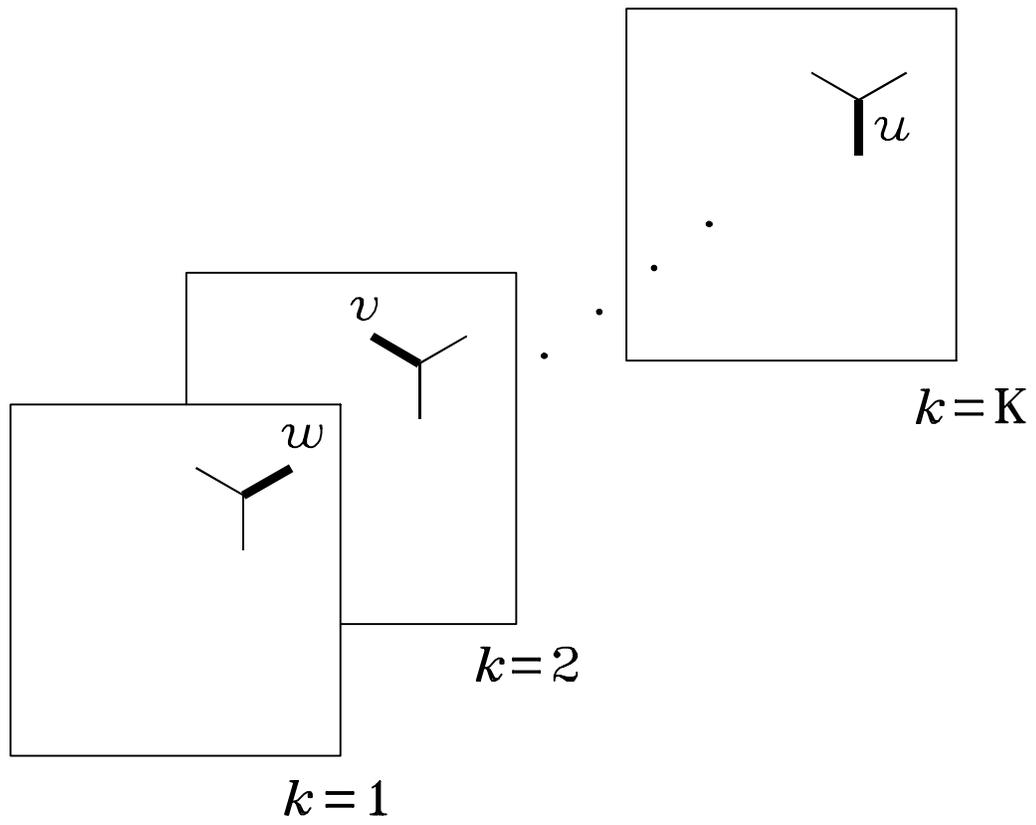
Fig. 2.  A stack of $K$ layers of honeycomb dimmer lattices.

\medskip
\noindent
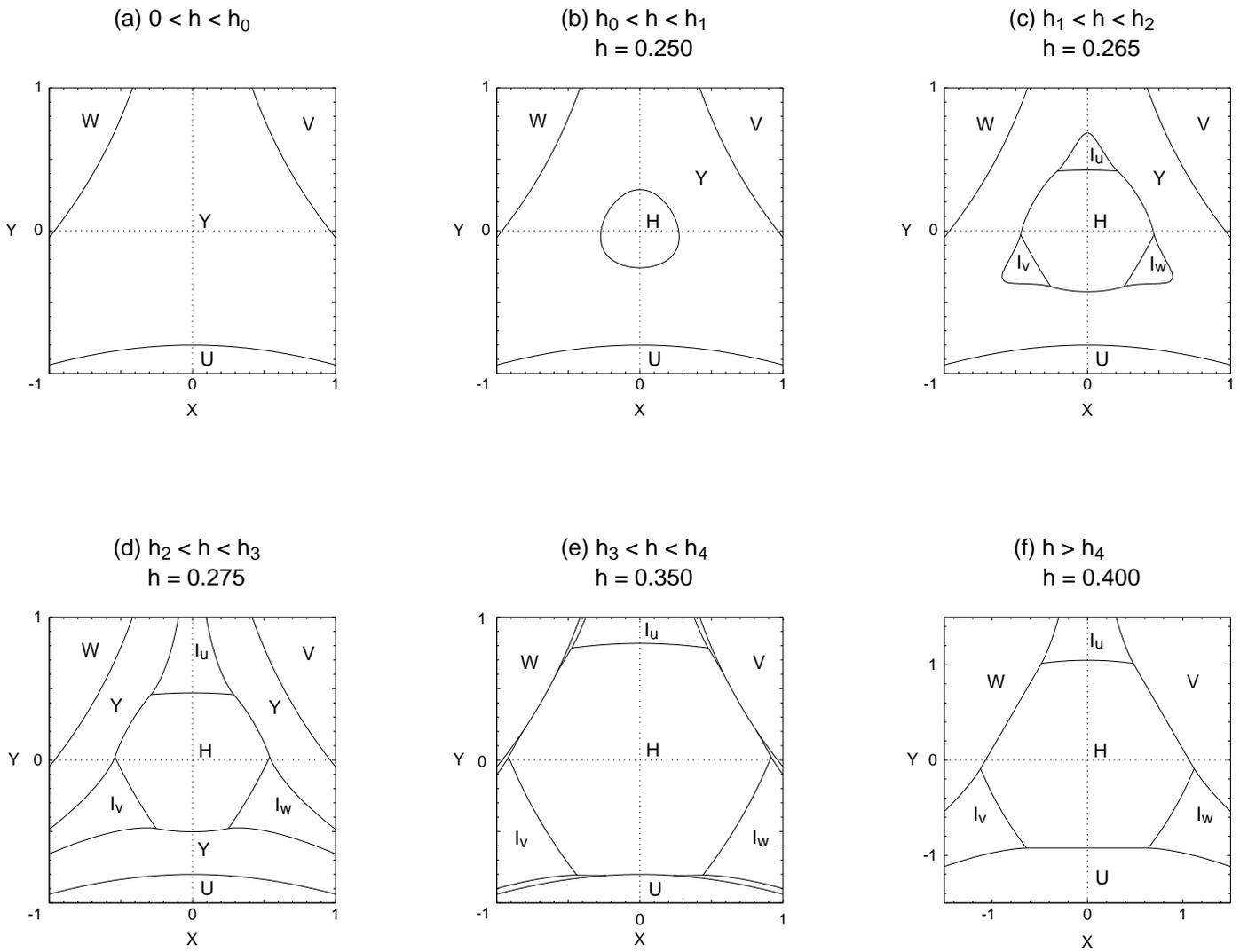
Fig. 3. The phase diagram.

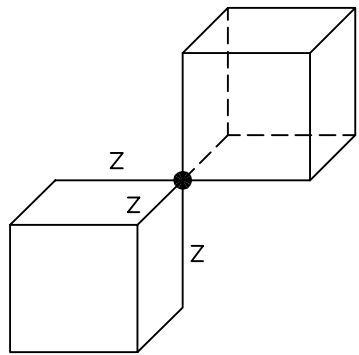
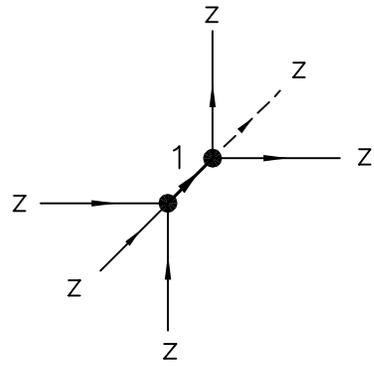
\end{document}